\documentstyle[12pt,amsfonts]{article}
\newcommand{\mbf}[1]{\mbox{\boldmath$ #1$}}

\newcommand{\be}{\begin{equation}}
\newcommand{\ee}{\end{equation}}
\newcommand{\ba}{\begin{eqnarray}}
\newcommand{\ea}{\end{eqnarray}}

\begin{document}

\begin{center}

{\LARGE\bf  Applications of the wave packet method to
resonant transmission and reflection gratings}

\vskip 1cm
{\Large Andrei G. Borisov ${}^{a,}$\footnote{email: {\sf borisov@lcam.u-psud.fr}} 
and  Sergei V. Shabanov ${}^{b,}$\footnote{email: {\sf shabanov@phys.ufl.edu}}}

\vskip 1cm

${}^a$ {\it Laboratory of Atomic and Molecular Collisions,
University of Paris-Sud, Orsay, France}

\vskip 0.3cm

${}^b$
{\it Department of Mathematics, University of Florida, Gainesville, FL 32611, USA}

\end{center}

\begin{abstract}
Scattering of femtosecond
laser pulses on resonant transmission and reflection gratings made
of dispersive (Drude metals) and dielectric materials is studied
by a time-domain numerical algorithm for  Maxwell's theory of
linear passive (dispersive and absorbing) media. The algorithm
is based on the Hamiltonian
formalism in the framework of which Maxwell's
equations for passive media are shown to be equivalent to
the first-order
equation, $\partial \Psi/\partial t = {\cal H}\Psi$, where 
${\cal H}$ is a linear differential operator (Hamiltonian) acting
on  a  multi-dimensional vector $\Psi$ built of the electromagnetic inductions
and auxiliary matter fields describing the medium response. The
initial value problem is then solved by means of 
a modified time leapfrog method in combination with the Fourier pseudospectral
method applied on a non-uniform grid that is constructed by 
a change of variables and designed to enhance the sampling efficiency
near medium interfaces. The algorithm is shown to be
highly accurate at relatively low computational costs.
An excellent agreement with
previous theoretical and experimental studies of the gratings
is demonstrated by numerical simulations using our algorithm.
In addition, our algorithm allows one to see real time dynamics
of long leaving resonant excitations of electromagnetic fields
in the gratings in the entire frequency range of the initial wide band  
wave packet as well as formation of the reflected and transmitted wave
fronts. 

\end{abstract}

\newpage

\section{Introduction}

The purpose of the present study is twofold. First, we test
a novel time-domain algorithm for the Maxwell's theory of linear,
passive (dispersive and absorbing) media. The algorithm is based
on ({\it i}) the Hamiltonian formalism for evolution differential 
equations \cite{arnold}, on ({\it ii}) the time
leapfrog scheme \cite{richt}, and on ({\it iii}) the Fourier pseudospectral
method \cite{fft} in combination with a change of variables that enhances
the spatial grid resolution in designated domains (in the vicinity
of medium interfaces) and, thereby, prevents the loss of accuracy
due to the aliasing problem
of the Fourier transform, while keeping the total spatial 
grid size fixed \cite{cc}, \cite{ab1}. Boundary conditions at medium interfaces are
not fixed in the algorithm, but rather medium parameters are 
allowed to have spatial discontinuities so that the correct
boundary conditions are enforced dynamically \cite{landau}, similarly to
the wave packet method for quantum mechanical systems with discontinuous
potentials.
Although, each of these three main ingredients
of our algorithm have been used individually in various computational
problems in electromagnetism and quantum mechanics, to our knowledge
they have never been put together in applications to numerical 
simulations of propagation of a wideband
electromagnetic pulse in passive media and its scattering on targets
made of dispersive and absorbing materials. 
By combining these methods, we have
obtained an efficient, true time-domain algorithm that is highly 
accurate, which is a known virtue of 
pseudospectral methods of solving partial differential equations \cite{psm}.

An essential advantage of time-domain numerical methods is that one can 
see all the immediate effects the medium and targets have on the
propagating wideband wave packet. Yet, a single simulation of
scattering of a
wideband wave packet is sufficient to determine some basic
electromagnetic properties of a target, e.g., transmission and
reflection coefficients, in the entire frequency range of the 
initial wave packet.

It has been reported \cite{gr1} that a periodic thin-film metallic
grating (either with holes or one-dimensional slits) 
can transmit more light at certain wavelengths than the
projected area of the holes (or slits) in the grating would suggest, while 
at other wavelengths transmission is almost fully blocked. There is
an ongoing discussion about the mechanism of such anomalous
transmission. Among suggested mechanisms are  formation of  dynamical diffraction
resonances in periodic metallic structures \cite{treacy}, surface
plasmons whose resonances are enhanced by the array of holes in the
grating \cite{gr2},
and, specific to the slit gratings, open Fabry-P\'erot resonant
cavities \cite{11a}. Since all the approaches produce essentially identical
predictions for the transmission and reflection coefficients
(the far-field) of the slit grating, it is, perhaps, necessary to
have a closer look at the details of the electromagnetic field
dynamics in the vicinity of the grating where deviations
of theoretical predictions of a particular mechanism from
the actual exact solution of Maxwell's equations might occur
(see, e.g., a discussion in a recent work \cite{11b}).
This is our second motivation of the present study. We apply our
algorithm to scattering of a wideband electromagnetic pulse on  
various transmission and reflection slit gratings (dispersive metallic and purely
dielectric ones). Our time-domain algorithm allows one to observe 
in details (in real time) formation and decay of long-living resonant excitations of
electromagnetic fields as well as formation of  resonant
transmission and reflection wave fronts. Thanks to the use of
the Fourier pseudospectral method in combination with nonuniform
grids, an extremely high accuracy of simulations can be achieved
in the entire simulation volume and time span at relatively
low computational costs.
We think that these virtues of our
time-domain algorithm would be useful for numerical studies of
electromagnetic properties of other nanostructured materials \cite{gr3}.

The paper is organized as follows.  Section 2 is devoted to the 
Hamiltonian formalism applied to the Maxwell's theory for 
general passive linear media. Maxwell's equations
for electromagnetic fields and medium responses are shown to
be equivalent to a first order evolution differential equation 
similar to the Schr\"odinger equation in quantum mechanics.
The wave function is a multidimensional column whose components
are electromagnetic fields and auxiliary fields describing 
the medium response. The Hamiltonian operator is a linear
differential operator acting on a Hilbert space spanned
by square integrable wave functions.
In Section 3, an example of the
multi-resonant Lorentz model, which is widely used to describe 
passive media, is considered 
in the framework of the Hamiltonian formalism. In 
Section 4, we establish a relation between the norm of 
wave functions and electromagnetic energy. The discussion is
limited to the Lorentz model and non-dispersive dielectric 
media. Section 5 is devoted to a general description of 
our algorithm. In particular, we show how the action of the  Hamiltonian
on wave functions is defined on a grid by means of the fast Fourier 
transform method. We prove that the Gauss law is enforced in the 
grid representation at no extra computational costs in our algorithm.
Then we discuss how the sampling efficiency of the fast Fourier method
can be enhanced in designated spatial regions
(typically at medium interfaces) by changing variables on the grid.
The time evolution is done by a modified leapfrog method applied
to the Schr\"odinger equation. We show that the conventional 
leapfrog method leads to an unstable algorithm for media with
absorption and propose a general method to modify the leapfrog scheme
to obtain a conditionally stable algorithm. Finally,
we give an explicit realization of our algorithm in the case of the
multi-resonant Lorentz model. Section 6 contains a detailed 
description of the actual computational scheme used in our simulations
of extraordinary transmission and reflection grating. We analyze the 
stability of the scheme using our general approach developed in
Section 5. Section 7 is devoted to our numerical results. We also
compare them with previous theoretical and experimental studies.
Section 8 contains a brief conclusion.

\section{Maxwell's theory for passive media in the Hamiltonian formalism}

Let ${\bf E}$ and ${\bf H}$ be electric and magnetic fields,
respectively, 
${\bf D}$ and ${\bf B}$ the corresponding inductions, and
${\bf P}$ and ${\bf M}$ the medium polarization and
magnetization vectors.  Boldface letters denote three-vector fields in ${\mathbb R}^3$.
Propagation of an electromagnetic wave packet in passive 
linear media in the absence of external radiating sources
is described by the following set of equations \cite{landau}
\ba
\label{1}
\dot{\psi}^I(t) &=& {\cal H}_0\psi^F(t)\ ,\\
\label{2} 
\psi^I(t)&=&\psi^F(t)+\psi^R(t)=\psi^F(t) +
\int_0^td\tau \chi(t-\tau )\psi^F(\tau)\ ,\\
\label{3}
\psi^I&=&\pmatrix{{\bf D}\cr{\bf B}}\ ,\ \ 
\psi^F=\pmatrix{{\bf E}\cr{\bf H}}\ ,\ \ 
\psi^R=\pmatrix{{\bf P}\cr{\bf M}}\ ,\ \ 
{\cal H}_0=\pmatrix{0&c\mbf{\nabla}\times\cr
-c\mbf{\nabla}\times & 0}\ ,
\ea
where  
$c$ is the speed of light in vacuum, the overdot denotes the partial derivative 
$\partial/\partial t$ with respect to time $t$, the
spatial argument of the fields, denoted below by ${\bf r}$, is 
suppressed. For generic anisotropic media, 
the medium response function $\chi(t)$ is regarded as
a linear operator (matrix) acting on the components of $\psi^F$.
For isotropic media, it is a scalar.
The response function is also position dependent
for non-homogeneous media.
Let $\psi^I(0)=\psi^F(0)$ be an initial wave packet with 
finite energy (finite ${\mathbb L}_2({\mathbb R}^3)$ norm).
We are interested in a finite norm solution of the initial value problem
for Maxwell equations (\ref{1}) subject to the constraint (the Gauss
law)
\be
\label{4}
\mbf{\nabla}\cdot{\bf B}(t)=\mbf{\nabla}\cdot{\bf D}(t)=0\ .
\ee

The response function $\chi(t)$ is usually deduced from a microscopic
model of the medium in question \cite{landau}. Therefore it is natural
to assume that $\chi(t)$ is a fundamental solution of some linear
evolution differential equation so that
\be
\label{5}
{\cal L}_t\psi^R(t) = \omega_p\psi^F(t)\ ,
\ee
where ${\cal L}_t$ is a linear differential operator (a polynomial in
$\partial/\partial t$) and $\omega_p$ 
describes a coupling between applied electromagnetic fields and
matter. In general, $\omega_g$ is a position dependent matrix.
Causality of the medium response requires that the Fourier transform
$\tilde{\chi}(\omega)$ 
of the response function should have poles only in the lower part
of the complex plane of $\omega$ \cite{landau} so that the Fourier
transform of (\ref{2}) reads 
$\tilde{\psi}^R(\omega)= \tilde{\chi}(\omega)
\tilde{\psi}^F(\omega)$. 
By taking the Fourier transform
of Eq. (\ref{5}), one finds that 
$\tilde{\cal L}(\omega) \tilde{\psi}^R(\omega) = 
\omega_p\tilde{\psi}^F(\omega)$ and, hence,
${\cal L}_t = \omega_p^{-1}[\tilde{\chi}(i\partial/\partial t)]^{-1}$.
If the response function is known from measurements in the frequency
domain, 
components of $[\tilde{\chi}(\omega)]^{-1}$ can always be approximated
by a polynomial with sufficient accuracy in a frequency
range of interest.

Now the Hamiltonian formalism \cite{arnold} can be applied to
(\ref{5}) to transform it to an equivalent system of first-order
differential equations
\be
\label{6}
\dot{\xi}(t) = {\cal H}_M^F\xi(t) + {\cal V}_{MF} \psi^F(t)\ ,
\ee  
where $\xi$ is a column of auxiliary fields which are linear combinations
of the response field and its time derivatives, save for the one of
the highest order. There exists a linear operator ${\cal R}$ such that 
$\psi^R(t)={\cal R}\xi(t)$. Its explicit form 
depends on the details of going over to the Hamiltonian
formalism. One can, for instance, identify the first component of
$\xi$ with $\psi^R$. In this case, ${\cal R}$ projects
the column $\xi$ onto its first component. The response function
can be expressed through 
the operators ${\cal H}_M^F$ and ${\cal V}_{MF}$ and ${\cal R}$
by making use of the fundamental solution of Eq. (\ref{6})
\be
\label{chi}
\chi(t) = \theta(t) {\cal R}e^{{\cal H}_M^Ft} {\cal V}_{MF}\ ,
\ee
where $\theta(t)$ is the Heaviside function. 
Equation (\ref{chi}) can be regarded as a condition on possible
choices of the operators ${\cal H}_M^F$ and ${\cal V}_{MF}$ and ${\cal R}$.

Applying ${\cal R}$ to (\ref{6}) we find 
\be
\label{7}
\dot{\psi}^R(t) = {\cal R}{\cal
  H}_M^F\xi(t)\ ,
\ee
 where the relation ${\cal R}{\cal V}_{MF}=0$ 
has been used. It is not hard to be convinced that the latter 
relation holds when the first component of $\xi$ coincides with
$\psi^R$. Any other choice of $\xi$ can be obtained by a canonical
transformation \cite{arnold} which is a linear nonsingular
transformation of auxiliary fields, 
\be
\label{ct}
\xi \rightarrow {\cal S}_M \xi\ ,\ \ \ \ \det{\cal S}_ M \neq 0\ .
\ee
According to (\ref{6}) and (\ref{chi}), ${\cal R}\rightarrow 
{\cal R}{\cal S}^{-1}_M$, ${\cal H}_M^F\rightarrow {\cal S}_M
{\cal H}_M^F{\cal S}^{-1}_M$, and ${\cal V}_{MF}
\rightarrow{\cal S}_M
{\cal V}_{MF}$ and, therefore, the condition ${\cal R}{\cal V}_{MF}=0$ remains
true in the new basis of auxiliary fields. Denoting ${\cal V}_{FM}=-{\cal R}{\cal H}_M^F$
and substituting (\ref{2}) and (\ref{7}) into (\ref{1})
Maxwell's evolution equations (\ref{1})  can be written in the Hamiltonian 
form
\be
\label{8}
\dot{\psi}^F(t) = {\cal H}_0\psi^F(t) + {\cal V}_{FM}\xi(t)\ .
\ee
Finally, the electromagnetic and auxiliary fields are unified into 
one column (wave function) so that Eqs. (\ref{8}) and (\ref{6})
can be represented as a single first-order evolution equation
\ba
\label{9}
\dot{\Psi}^F(t) &=&{\cal H}^F\Psi^F(t)\ ,\\
\label{10}
\Psi^F&=&\pmatrix{\psi^F\cr \xi}\ ,\ \ {\cal H}^F=\pmatrix{
{\cal H}_0 &{\cal V}_{FM}\cr {\cal V}_{MF}&{\cal H}_M^F}\ .
\ea
The index $F$ indicates that electromagnetic fields are used
as independent electromagnetic degrees of freedom. We shall refer
to (\ref{10}) as to
 a field representation. Accordingly, an induction representation
is obtained by the similarity transformation
\be
\label{11}
\Psi^I=\pmatrix{\psi^I\cr \xi}={\cal S}\Psi^F\ ,\ \ \ 
{\cal S}=\pmatrix{1&{\cal R}\cr 0&1}\ ,\ \ 
{\cal H}^I = {\cal S}{\cal H}^F{\cal S}^{-1}\ .
\ee
The corresponding blocks of ${\cal H}^{I}$ have the form
\ba
\label{IF1}
{\cal H}_I &=& {\cal H}_0 \ ,\ \ \ \ \ \ \ \ \ {\cal V}_{MI} = {\cal V}_{MF}\ ,\\
\label{VIM}
{\cal V}_{IM} &=& {\cal V}_{FM} + 
{\cal R}{\cal H}_M^F - {\cal H}_0{\cal R} = -{\cal H}_0{\cal R}\ ,\\ \label{HI}
{\cal H}^{I}_M &=& {\cal H}^{F}_M - {\cal V}_{MF}{\cal R}\ .
\ea
In what follows we denote wave functions by $\Psi^Q(t)$ with $Q$ being
the representation index, $F$ or $I$.
This completes construction of the Hamiltonian representation 
of Maxwell's theory for linear passive media.

Boundary conditions
at medium interfaces and possible scatterers (targets)
are not imposed on electromagnetic fields, but rather
they are enforced dynamically by allowing medium parameters
to be discontinuous functions.
The fundamental solution of (\ref{9}) 
\be
\label{12}
\Psi^F(t) = e^{t{\cal H}^F}\Psi^F(0)\ ,\ \ \ t\geq 0\ ,
\ee
is well defined for discontinuous ``potentials'' ${\cal V}_{MF}$
and ${\cal V}_{FM}$, for example, by means of the
Kato-Trotter product formula used in the path integral representation
of (\ref{12}) as shown in \cite{maxwell}.

\section{The Lorentz model}
\setcounter{equation}0

The Hamiltonian formalism for the Lorentz model has been used in \cite{hflm} to develop 
a finite differencing algorithm to study an electromagnetic
pulse propagation in Lorentz media. Here we derive an explicit form 
of the Hamiltonian for the Lorentz model which is used in Section 5
for the stability analysis of our algorithm. 
The Lorentz model of a passive medium is based on the 
assumption that the medium magnetization is zero, ${\bf M}=0$, while 
the medium polarization is described by a set of decoupled 
second-order differential equations \cite{landau}
\be
\label{p}
\ddot{\bf P}_a + 2\gamma_a\dot{\bf P}_a +
\omega_a^2 {\bf P}_a = \omega_{pa}^2 {\bf E}\ , \ \ \ \ \ 
{\bf P} = \sum_{a=1}^N {\bf P}_a\ .
\ee
where $\omega_a$ are resonant frequencies, $\gamma_a$ are
damping coefficients, and $\omega_{pa}$ are plasma frequencies.
As has been pointed out,
no boundary conditions are imposed on electromagnetic
fields at medium and/or target interfaces. 
Instead, the coupling constants $\omega_{pa}=
\omega_{pa}({\bf r})$ are allowed to have discontinuities at medium
interfaces, or,
from the physical point of view, they remain smooth but change
rapidly, $\lambda_w|\mbf{\nabla }\omega_p|/\omega_p >\!\! > 1$,
at the interface, where $\lambda_w$ is a typical wave length
of the incident wave packet. The initial
value problem is solved in the space of square integrable 
wave functions. Initial conditions for 
the response field are
${\bf P}(t=0)=\dot{\bf P}(t=0)=0$.

Consider $2N$ real vector fields, $\mbf{\xi}_j$, 
$j=1,2,...,2N$, such that
\ba
\label{aux}
{\bf P}_a &=&(\omega_{pa}/\omega_a)\, \mbf{\xi}_{2a-1} \ ,\\
\label{aux1}
\dot{\mbf{\xi}}_{2a-1} &=& \omega_a\mbf{\xi}_{2a}\ ,\ \ \ \ 
\dot{\mbf{\xi}}_{2a}  
= -2\gamma_a\mbf{\xi}_{2a} -\omega_a\mbf{\xi}_{2a-1} + \omega_{pa}{\bf E}\ .
\ea
Thus, the original system of second order equations has been
converted into the first order system. The operator ${\cal R}$ 
is defined by (\ref{aux}).
After simple algebraic transformations, we infer
\ba
\label{lmvfm}
{\cal V}_{FM} &=& ({\cal V}_{FM1}, {\cal V}_{FM2}, 
\cdots, {\cal V}_{FMN})\ ,\ \ \ \ {\cal V}_{FMa} = 
\pmatrix{0 & -\omega_{pa}  \cr 0&0}\ ,\\ \label{lmvmf}
{\cal V}_{MF} &=&- {\cal V}_{FM}^\dagger\ ,\\ \label{lmhm}
{\cal H}_M^{F} &=&{\sf diag\ }\left({\cal H}_{M1}^{F},\ {\cal
H}_{M2}^{F}, 
\cdots, {\cal H}_{MN}^{F}\right)\ , \ \ \ \ 
{\cal H}_{Ma}^{F} = \pmatrix{0 & \omega_a \cr -\omega_a & -2\gamma_a}\ ,
\ea
where ${\sf diag}$ indicates that the corresponding matrix 
is block-diagonal with blocks listed in the order 
from the upper left to lower right corners. Note that the matrices
${\cal V}_{FMa}$ and ${\cal H}_{FMa}$ act on a six-dimensional column
$\xi_a$ composed of two vectors $\mbf{\xi}_{2a-1}$ and $\mbf{\xi}_{2a}$. 
Therefore they should be 
understood as composed of $3\times 3$ blocks. Each block is obtained
by multiplying the unit matrix by the number indicated in place of the
block in (\ref{lmvfm}) and (\ref{lmhm}).

Another convenient way to introduce 
the Hamiltonian formalism is to use $N$ complex
vector fields $\mbf{\zeta}_a$ which satisfy the first order differential equation
\be
\label{caux}
\dot{\mbf{\zeta}}_a = \lambda_a \mbf{\zeta}_a -
i\omega_{pa}{\bf E}\ ,\ \ \ \
{\bf P}_a = \frac{\omega_{pa}}{2\nu_a}\ 
\left(\mbf{\zeta}_a + \bar{\mbf{\zeta}}_a\right)\ ,
\ee
where $\lambda_a = -\gamma_a + i\nu_a$ and $\nu_a = \sqrt{\omega_a^2 -\gamma_a^2}$. 
This representation is defined only if $\gamma_a < \omega_a$ (i.e., the attenuation is
not high). From the numerical
point of view, solving a {\it decoupled} system of 
$N$ first order differential equation and taking complex conjugation 
(denoted here by an over bar)  is less expensive 
than solving the original system of second-order differential equations
for the medium polarization. 
In general, there is always a freedom
of choosing a new basis for the auxiliary field space (\ref{ct}). 
If the evolution operator $
\exp(t {\cal H}^{Q})$ 
is computed in one basis, it can be computed in another
basis by a suitable
similarity transformation. This is an important observation because
the auxiliary field 
basis can be chosen  in a way that
facilitates computation of the evolution operator, e.g., to 
speed up simulations. For instance,
in the complex representation (\ref{caux}), 
the matter Hamiltonian ${\cal H}_M^F$ is
diagonal. The corresponding transformation of auxiliary fields is given by
\be
\label{S}
\pmatrix{\mbf{\xi}_{2a-1}\cr \mbf{\xi}_{2a}} = \frac{1}{2\nu_a}\, 
\pmatrix{\omega_a& \omega_a\cr \lambda_a &\bar{\lambda}_a}
\pmatrix{\mbf{\zeta}_a \cr \bar{\mbf{\zeta}}_a} 
\equiv {\cal S}_M \pmatrix{\mbf{\zeta}_a \cr \bar{\mbf{\zeta}}_a}\ .
\ee
To transform the whole system into this representation, the Hamiltonian ${\cal H}^{F}$ is 
replaced by ${\cal S}^{-1}{\cal H}^{F}{\cal S}$ and the wave function $\Psi^F$ by
${\cal S}\Psi^F$ where ${\cal S}$ is 
block-diagonal with the unit matrix in the upper left (field) corner
and with ${\cal S}_M$ in the lower right (matter) corner. 

\section{Energy and the norm of state vectors}
\setcounter{equation}0

Accuracy and convergence of a numerical algorithm is
defined relative to some norm.
Let us discuss the choice of a norm in the space
spanned by wave functions $\Psi^Q(t)$. The discussion
is limited to the Lorentz model and the case of a nonhomogeneous,
nondispersive medium (dielectric) which are used in our numerical
simulations. 

\subsection{The Lorentz model}

Consider a
multi-resonant Lorentz model with no attenuation $\gamma_a=0$.
The field and matter evolution
equations can be obtained from the variational principle for the action
\be
\label{action}
S = \int dt L = \int\! dt\!\!\int\! d{\bf r}\left[
\frac 12 \left( {\bf E}^2 - {\bf B}^2\right)
+ \frac 12 \sum_a \left( \dot{\mbf{\vartheta}}_a^2 -
\omega_a^2\mbf{\vartheta}^2_a\right)
+ {\bf P}\cdot {\bf E}\right]\ ,
\ee
where the polarization of the medium is expressed via matter fields as
${\bf P}= \sum_a \omega_{pa} \mbf{\vartheta}_a$.
The electromagnetic degrees of freedom are described by the vector
and scalar potentials, respectively, ${\bf A}$ and $\varphi$.
The fields are defined by 
${\bf E}= -\mbf{\nabla} \varphi -
\dot{\bf A}$ and ${\bf B}= \mbf{\nabla}\times {\bf A}$.
The units are chosen in this Section so that $c=1$.
The least action principle for the scalar potential $\varphi$ leads 
to the Gauss law, $\mbf{\nabla}\cdot {\bf D} = 0$, for the vector
potential ${\bf A}$ to the Maxwell's equation, $\dot{\bf D} 
= \mbf{\nabla}\times {\bf B}$, and for the matter fields
$\mbf{\vartheta}_a$ to the medium polarization evolution equation
(\ref{p}) 
of the  Lorentz model with no attenuation, $\gamma_a=0$. 
The second Maxwell's equation and the Gauss law
for the magnetic field follows from the relation 
${\bf B}= \mbf{\nabla}\times {\bf A}$ by taking its time
derivative and divergence, respectively. 

The energy of the system
coincides with the canonical Hamiltonian which is obtained by
the Legendre transformation \cite{arnold} 
of the Lagrangian $L$ for the velocities $\dot{\bf A}$
and $\dot{\mbf{\vartheta}}_a$.  
The canonical momenta are $\mbf{\pi}_a = \delta L/\delta
\dot{\mbf{\vartheta}}_a = \dot{\mbf{\vartheta}}_a$
and $\mbf{\Pi} = \delta L/\delta \dot{\bf A} = -{\bf E} -{\bf P}=
-{\bf D}$. Doing the Legendre transformation,
we find the canonical Hamiltonian (energy) of the system
\be
\label{energy}
{E}(t) = \int\! d{\bf r}\left(\sum_a\mbf{\pi}_a\cdot\dot{\mbf
{\vartheta}}_a +\mbf{\Pi}\cdot\dot{\bf A}\right) -L=
\frac 12 \int\! d{\bf r} \left[
{\bf E}^2 + {\bf B}^2 + \sum_a\left(
\mbf{\pi}^{2}_a + \omega_a^2 \mbf{\vartheta}^{2}_a\right)\right] \ ,
\ee
where the Gauss law $\mbf{\nabla}\cdot {\bf D}=0$ has been used. 
The energy conservation
follows directly from the Noether theorem \cite{arnold} applied to
the time translation symmetry of the action (\ref{action}),
$\dot{E}(t)=0$.  Equation (\ref{energy}) becomes the conventional expression for
the electromagnetic energy in a passive medium \cite{landau} when $\mbf{\pi}_a$
and $\mbf{\vartheta}_a$ are replaced by the corresponding 
solutions of the equations
of motion with initial conditions $\mbf{\pi}_a(t=0)=\mbf{\vartheta}_a(t=0)=0$. 

An important observation is that
the Noether integral of motion (\ref{energy}) coincides with
the norm squared of the corresponding state vector 
\be
\label{enorm}
E = \frac 12 \int d{\bf r}\, \Psi^{F\dagger}\Psi^F \equiv
(\Psi^F,\Psi^F)=
(\Psi^I, \mu \Psi^I)\ ,
\ee
where $\mu = {\cal S}^{-1\dagger}{\cal S}^{-1}$. This follows from the 
fact that, if we identify
$\mbf{\xi}_{2a} = \mbf{\pi}_a$
and $\mbf{\xi}_{2a-1} = \omega_a\mbf{\vartheta}_a$, the Hamiltonian equations of motion,
$\dot{\mbf{\vartheta}}_a=\delta E/\delta \mbf{\pi}_a$ 
and $\dot{\mbf{\pi}}_a=-\delta E/\delta \mbf{\vartheta}_a$,
coincide with  (\ref{aux1}) when $\gamma_a=0$.  
Note that canonically conjugated electromagnetic variables 
are ${\bf A}$ and  $-{\bf D}$. Therefore, 
the coupling between the electromagnetic and matter degrees of
freedom in the Hamiltonian equations of motion is generated by the term 
${\bf E}^2 = ({\bf D} -{\bf P})^2$ in (\ref{energy}).
Thus, in the absence of attenuation, the norm of the state
vector is proportional to the wave packet electromagnetic energy
and, hence, is conserved. 

The norm conservation also follows 
from anti-Hermiticity of the Hamiltonian 
${\cal H}^{F\dagger}=-{\cal H}^F$ if $\gamma_a=0$, while (\ref{enorm})
establishes a relation between the electromagnetic energy and the 
norm. In the induction representation, the norm in the measure space,
defined by the operator $\mu$ in (\ref{enorm}),
is also conserved by construction. Consequently,
the Hamiltonian is anti-Hermitian relative to the measure space
scalar product, ${\cal H}^{I\dagger}\mu = -\mu {\cal H}^I$.

The norm (energy) conservation can be used to control numerical 
convergence, especially when the aliasing problem
in the fast Fourier transform is present, e.g., 
when parameters of the medium are
discontinuous functions in space. In a properly designed algorithm
the loss of energy (norm) due to attenuation should be
controlled by the symmetric part of the Hamiltonian operator
\be
\label{attoper}
\dot{E}(t) = -\sum_a \gamma_a \int\! d{\bf r}\ \mbf{\xi}_{2a}^2(t) \equiv
\frac 12\, \left(\Psi^F(t), {\cal V}_\gamma^F\Psi^F(t)\right) \leq 0\ ,
\ee
where ${\cal V}_\gamma^{F\dagger} = {\cal V}_\gamma^F 
= ({\cal H}^{F\dagger} 
+ {\cal H}^F)/2 \leq 0$ (a negative semidefinite operator) 
which is, in this case, a diagonal matrix with nonpositive
elements. 

\subsection{Nondispersive media}

If the medium in question does not have dispersion and absorption,
the formalism is simplified. 
Let $\varepsilon = \varepsilon({\bf x})$ be the dielectric
constant of the medium, ${\bf D}=\varepsilon{\bf E}$. If the medium is not isotropic,
then $\varepsilon$ is symmetric positive definite $3\times 3$
matrix everywhere in space.
We rewrite Maxwell's equations in the form
\be
\label{schr}
\dot{\psi}^I(t) = {\cal H}_G\psi^I(t)\ ,\ \ \ \ {\cal H}_G =  
\pmatrix{0 & c\mbf{\nabla}\times \cr -
c\mbf{\nabla}\times (\varepsilon^{-1}\  \ )& 0\cr} \ ,
\ee
where the parenthesis in $(\varepsilon^{-1})$ 
mean that the induction is first multiplied by $\varepsilon^{-1}$
and then the curl of the resulting vector field is computed. 
Consider the scalar product
\be
\label{sp}
(\psi_1^I,\mu_\varepsilon\psi_2^I) = 
\int d{\bf r}\ \psi^{I\dagger}_1\mu_\varepsilon \psi_2^I\ ,\ \ \ 
\mu_\varepsilon = 
\pmatrix{\varepsilon^{-1} & 0 \cr 0& 1\cr}\ .
\ee 
The Hamiltonian is anti-Hermitian with respect to this measure
space scalar product,
${\cal H}^\dagger_G\mu_\varepsilon = -\mu_\varepsilon{\cal H}_G$.
Therefore the corresponding norm is preserved in the time evolution
generated by $\exp(t{\cal H}_G)$, that is, 
$(\psi^I(t),\mu_\varepsilon\psi^I(t)) = 
(\psi^I(0),\mu_\varepsilon\psi^I(0))$. The electromagnetic energy of the wave packet
is conserved because it is proportional to the measure space norm of the 
initial state vector.

\section{The algorithm}
\setcounter{equation}0

\subsection{The grid representation}

Consider an equidistantly spaced finite grid with periodic 
boundary conditions. Let $\Delta r$ be
the grid step and ${\bf n}$ be a vector with integer valued
components. Then the dynamical variables are projected onto the grid
by taking their values at grid points ${\bf r}= {\bf n }\Delta r$,
that is, the wave function is replaced by a vector (column)
whose dimension is determined by the grid size,
$\Psi^Q({\bf r}) \rightarrow \Psi^Q({\bf n}\Delta r)\equiv\Psi^Q_{\bf n}$.
A cubic grid is assumed.
Consider a discrete
Fourier transformation associated with the grid, 
$
\tilde{\Psi}^Q ({\bf n}k_0) = \sum_{{\bf n}^\prime}{\cal F}_{{\bf
nn}^\prime} \Psi^Q_{{\bf n}^\prime}\equiv \tilde{\Psi}_{\bf n}^Q$, 
where $ {\cal F}^\dagger{\cal F}= {\cal
F}{\cal F}^\dagger = 1$.
The reciprocal lattice step is $k_0= 2\pi/\Delta r$. The grid spatial size
$L$ and step $\Delta r$ must be chosen so that the Fourier transform of the initial
wave packet has support within the region
$k\in [k_{min}, k_{max}]$ where $k=|{\bf k}|$,  $k_{max} = k_0$ and $k_{min} = 2\pi/L$.
The action is of any position dependent operator ${\cal V}^Q={\cal
  V}^Q({\bf r})$ is defined by
\be
\label{grid4}
{\cal V}^Q({\bf r})\Psi_t^Q({\bf r})
\rightarrow {\cal V}^Q({\bf n}\Delta r) \Psi_t^Q({\bf n}\Delta
r)\equiv
\sum_{{\bf n}^\prime} {\cal V}^Q_{{\bf n}{\bf n}^\prime}\Psi_{{\bf n}^\prime}^Q \ ,
\ee
where ${\cal V}^Q_{{\bf n}{\bf n}^\prime}=\delta_{{\bf n}{\bf
    n}^\prime}
{\cal V}^Q({\bf n}\Delta r)$ is a diagonal matrix.
Let ${\cal H}^Q_0={\cal H}_0^Q(\mbf{\nabla})$ depend only
on the ${\mbf{\nabla}}$ operator.  The
projection of its action onto the grid is then 
defined via the discrete Fourier transform
\be
\label{grid5}
\left.
{\cal H}_0^Q(\mbf{\nabla}) \Psi_t^Q({\bf r})\right|_{{\bf r}= {\bf
n}\Delta r}
\rightarrow\sum_{{\bf n}^\prime} {\cal H}^Q_{0{\bf n}{\bf n}^\prime}
\Psi_{{\bf n}^\prime}^Q \ ,\ \ \ \ 
{\cal H}^Q_{0{\bf n}{\bf n}^\prime} = \sum_{{\bf n}^{\prime\prime}}
\left({\cal F}^\dagger\right)_{{\bf
n}{\bf n}^{\prime\prime}} {\cal H}_0^Q(i{\bf n}^{\prime\prime} k_0)
\left({\cal F}\right)_{{\bf
n^{\prime\prime}}{\bf n}^{\prime}}\ .
\ee
The projection (\ref{grid5}) is performed by the fast Fourier transform method. 

In what follows, the action of a product of ${\cal V}^Q$ and ${\cal H}^Q_0$
on any state vector is understood as multiplication of $\Psi_{\bf
  n}^Q$ by the corresponding matrices, defined in  
(\ref{grid4}) and (\ref{grid5}),
in the order specified in the product. The main advantage of using
the Fourier basis is the exponential convergence (versus 
the polynomial one in finite differencing schemes) \cite{psm}
as the grid size increases, which allows one to substantially
increase the accuracy of simulations.

\subsection{The Gauss law}

Another advantage of the Fourier basis is that the Gauss law 
is enforced at no extra computational cost. 
In the grid representation defined above, the Gauss law
(\ref{4}) requires that the Fourier transforms
of the inductions $\tilde{\bf D}({\bf k})$ and $\tilde{\bf B}({\bf k})$
remains perpendicular to the reciprocal grid vector ${\bf k}=
{\bf n}k_0$ at any moment of time. In our algorithm, as we shall
show shortly, the time evolution is generated by applying 
powers of the Hamiltonian to the wave function. In the induction
representation, the action of powers of ${\cal H}^I$ always 
produces the cross product ${\bf k}\times \tilde{\bf C}({\bf k})$,
for some $\tilde{\bf C}({\bf k})$ regular at ${\bf k}=0$, in the entries of
$\Psi^I$ that correspond to the electromagnetic inductions.
Hence, in the grid representation the wave function 
$({\cal H}^I)^m\Psi^I$ satisfies the Gauss law for any power $m$
because of the trivial identity ${\bf k}\cdot ({\bf k}\times 
\tilde{\bf C}({\bf k}))=0$ valid for any vector ${\bf k}$ 
of the reciprocal grid. 

It should be noted that a high accuracy of the Gauss law is
essential to achieve a high accuracy of simulated electromagnetic
fields near medium interfaces. 

\subsection{Improving sampling efficiency by changing variables}

As is well known from the Fourier analysis, the convergence rate 
can be affected for functions which 
have discontinuities \cite{fft}. The latter is, unfortunately, the case
in electromagnetic scattering problems \cite{landau}. Suppose there is an interface
between two media. It can be deduced from 
the dynamical Maxwell's equations that the components of 
electric and magnetic fields, ${\bf E}$ and ${\bf H}$,
tangential to the interface must be continuous, provided there is
no surface electric current on the interface. From the Gauss law
it follows that components of the inductions, ${\bf D}$
and ${\bf B}$, normal to the interface must be continuous, provided
there is no surface charge on the interface. In contrast, 
normal components
of the fields and tangential components of the inductions can be
discontinuous. Their discontinuities are determined by 
discontinuities of medium parameters (e.g., discontinuities in
plasma frequencies in Lorentz models).
Therefore, in either the
induction or field representation, there are components which suffer
discontinuities at the interface. 
The only way to cope with the problem, while keeping the use
of the Fourier basis, is to make the grid finer \cite{fft}. This would 
lead to a substantial waste of computational resources because
the conventional fast Fourier transform is defined on a uniform
periodic grid, while 
the sampling efficiency should only be enhanced in the neighborhood
of medium interfaces. The use of wavelet bases might be helpful
for such a task \cite{daubl} in time domain algorithms. 
Here we retain the Fourier basis,
and increase the sampling efficiency by a change of variables.

The basic idea can be understood with a one-dimensional example.
Let $z$ be a physical coordinate. Consider a change of variables
defined by $z=f(y)$ where $y$ is an auxiliary coordinate.
An equidistant grid $y_n=n\Delta y$, with $n$ being integers,
of the auxiliary coordinate generates a non-uniform grid of the
the physical coordinate, $z_n = f(n\Delta y)$. Assuming 
$\Delta y$ to be sufficiently small and $f(y)$ sufficiently
smooth, the physical grid spacing can be approximated as
$$
\Delta z_n = z_{n+1}-z_n \approx \Delta y f^\prime (n\Delta y)\ .
$$ 
So, if $f^\prime(y)=1$, then $\Delta z_n =\Delta y$ and
the grid is equidistant.  
By making the derivative $0<f^\prime(y)\leq 1$ in some designated
areas, one can achieve a desired local grid density in 
the physical space, while keeping the total
grid size fixed. For example, if it is necessary to
increase the sampling efficiency in the vicinity $z=0$,
one can take $f^\prime(y) = 1- a_0[1+b_0^2y^2]^{-1}$,
where $0<a_0<1$, and , hence,
$$
f(y) = y -\frac{a}{b}\tan^{-1}(by)\equiv y-g(y,a,b)\ .
$$
By adjusting parameters $a$ and $b$, the local grid 
density can be changed as desired. Consequently,
if the sampling efficiency is to be enhanced at
several points $y_i$, a suitable change of variables
can be of the form $f(y)=y-\sum_i g(y-y_i,a_i,b_i)$.  
The fast Fourier transform is applied on a uniform grid
of the auxiliary variable.

Upon the change of variables, the integration measure
in the scalar product and the derivative transform,
respectively, $dz = dy f^\prime(y)$ and $\partial_z
=[f^\prime(y)]^{-1}\partial_y$. When projected on
a uniform grid of the physical coordinate by means of
(\ref{grid5}), the derivative
operator $\partial_z$ becomes an anti-Hermitian matrix.
When the rule (\ref{grid5}) is applied on a uniform
grid of the auxiliary coordinate $y$, the derivative
operator $\partial_z$ is no longer represented by
an anti-Hermitian matrix, although it is still 
an anti-Hermitian linear operator, but in the 
measure space where the scalar product is defined
with the weight $f^\prime(n\Delta y)$ at each
lattice cite. From the numerical point of view
it is convenient to have an explicitly anti-Hermitian
matrix representation of the derivative on the grid.
Due to round-off errors, the exact anti-Hermiticity
of $\partial_z$ in the measure space can be violated
in the grid representation, which, in turn, may lead 
to numerical instabilities of simulations. 
Note that if a computed matrix is known to be anti-Hermitian,
then only half of its elements is to be computed, while
the other elements are restored by the symmetry. In the 
measure space, an explicit anti-Hermiticity of a linear operator 
is much more difficult to maintain in numerical simulations 
because the symmetry relation between matrix elements depends
on the scalar product measure. 
For this reason, the wave function is rescaled by the square
 root of the Jacobian \cite{ab1}, $\Psi \rightarrow \sqrt{f^\prime}
\Psi$ so that the integration measure becomes $dy$ leading
to the conventional Euclidean scalar product 
in the grid representation (with the uniform unit
weight at each grid site). In such a representation,
the derivative operator $\partial_z \rightarrow
[f^\prime]^{-1/2}\partial_y[f^\prime]^{-1/2}$
becomes again an explicitly anti-Hermitian
matrix in the grid representation defined
by the rules (\ref{grid4}) and (\ref{grid5}).

\subsection{A modified temporal leapfrog scheme}

A conventional temporal leapfrog method of solving
the Schr\"odinger equation is based on the iteration
scheme \cite{richt}
\be
\label{lf1}
\Psi^Q(t+\Delta t) = \Psi^Q(t-\Delta t) + 2\Delta t {\cal H}^Q \Psi^Q(t)\ ,
\ee
so that the wave function in the consecutive time moment is computed
from the wave function at two previous moments of time, where $\Delta t$
is the time step. The action of the Hamiltonian ${\cal H}^Q$ is
computed by pseudospectral methods, in particular, by means of the Fourier basis
and the fast Fourier transform in our algorithm. The leapfrog algorithm
is conditionally stable for an anti-Hermitian ${\cal H}^Q$, which is
true for nondispersive media, but not the case for media with absorption.  

To investigate the stability,
let us introduce the amplification matrix for the leapfrog
algorithm, $\Psi(t+\Delta t) = {\cal G}(\Delta t)\Psi(t)$
(the representation index, $Q$, is suppressed for a moment).
From (\ref{lf1}) it follows that ${\cal G}(\Delta t)$
satisfies the equation ${\cal G}^2(\Delta t) - 2\Delta t {\cal H}
{\cal G}(\Delta t) -1=0$ which has two solutions 
\be
\label{l2}
{\cal G}^{(\pm)}(\Delta t) = {\cal H}\Delta t \pm 
\sqrt{1 +{\cal H}^2\Delta t^2} \ .
\ee
The stability of the algorithm requires that the energy norm
of both approximate solutions $[{\cal G}^{(\pm)}(\Delta t)]^n\Psi(0)$
must be uniformly bounded for $n>0$. The necessary condition (but not sufficient) 
is the von Neumann condition that the spectral radius of the 
amplification matrix does not exceed 1. If a complex number 
$Re^{i\varphi}$
is an eigenvalue of $\Delta t {\cal H}$, then the von Neumann
condition implies that  $\varphi =\pm \pi/2$ and $R^2 \leq 1$.
In other words, eigenvalues of $\Delta t {\cal H}$ must be imaginary and their
magnitude should not exceed 1. If in addition we demand that
the Hamiltonian is diagonalizable, then 
a conditional stability can be achieved for sufficiently small
$\Delta t$. Indeed, in this case
there exists an non-singular ${\cal S}$ such that
\be
\label{stability}
{\cal H} = {\cal S}^{-1}{\cal H}_S{\cal S}\ ,\ \ \ \ 
{\cal H}^{\dagger}_S= -{\cal H}_S\ ,
\ee
and $\Delta t{\cal H}_S$ satisfies the von Neumann condition. 
Since ${\cal H}$ and ${\cal H}_S$ have the same 
eigenvalues, the amplification matrices for  
${\cal H}$ and ${\cal H}_S$ are related by the same similarity
transformation (\ref{stability}). Hence, the norm of 
a wave function obtained by  
the action of powers of (\ref{l2}) on
an initial wave function is uniformly bounded.
Note also that the Hamiltonian (\ref{stability}) is anti-Hermitian
relative to the measure space scalar product, 
${\cal H}^\dagger\mu=-\mu{\cal H}$, where 
$\mu={\cal S}^{-1\dagger}{\cal S}^{-1}$. The stability can also be
proved via the equivalence of the conventional Euclidean norm 
and the $\mu$-norm \cite{maxwell}.

In the case of nondispersive media, the Hamiltonian ${\cal H}_G$ is anti-Hermitian
in the measure space, and the von Neumann condition is fulfilled if
\be
\label{scdiel}
\Delta t\, c\, k_{max}^\varepsilon\leq  1 \ ,
\ee
where $k_{max}^\varepsilon$ 
the maximal norm of all wave vectors in the medium which
can be estimated by $\sqrt{\rho(\varepsilon)}k_{max}$
with $k_{max}$ being the maximal wave vector of the initial
pulse in vacuum and
$\rho(\varepsilon)$ the maximal spectral
radius of the symmetric matrix $\varepsilon({\bf x})$ over ${\bf x}$ (or simply
the maximum of $\varepsilon({\bf x})$ if the medium
is isotropic). This can be understood from the following
principle \cite{shin}. A finite difference scheme with variable
coefficients is stable if all the corresponding schemes
with frozen (i.e., fixed to a particular value everywhere
in space) coefficients are
stable. 

The von Neumann condition cannot be met 
if absorption is present because 
eigenvalues of the Hamiltonian must have real parts in order to
account for exponential attenuation of field amplitudes. To circumvent this
difficulty, the leapfrog scheme is modified in the following
way \cite{maxwell}. We assume the Hamiltonian to be
diagonalizable. The lack of eigenvectors of the Hamiltonian
typically leads to solutions whose amplitudes grow 
polynomially in time \cite{maxwell}.
This feature cannot be present in a physically reasonable model
of passive media. So our assumption is justified from the 
physical point of view and, yet, the Lorentz model Hamiltonian is
indeed diagonalizable. Let
\be
\label{split}
{\cal H}= {\cal H}_0 + {\cal V}\ ,\ \ \ \ (\Psi,{\cal V}\Psi)\leq 0
\ee
for any $\Psi$ in the Hilbert space and the von Neumann condition 
is satisfied  for ${\cal H}_0$, i.e., ${\cal H}_0$ has imaginary
eigenvalues. The split (\ref{split}) can be achieved in many ways.
For instance, ${\cal H}_0$ can be obtained from ${\cal H}$ by 
setting all parameters responsible for attenuation to zero.
In a physically acceptable model, the energy must be conservative
in an absorption free medium and so must be the energy norm
of wave vectors, and, therefore, the corresponding Hamiltonian
must be anti-Hermitian (relative to the energy induced scalar product).
In the Lorentz model this is easily seen in the field 
representation if we set ${\cal H}_0 = 
{\cal H}\vert_{\gamma_a=0}$ which is explicitly anti-Hermitian
(see Section 3).
Then ${\cal V}$ is diagonal with
matrix elements being zeros and $-2\gamma_a$. 
Another possibility
is to identify ${\cal H}_0$ with the Hamiltonian in the vacuum,
then ${\cal V}={\cal H}-{\cal H}_0$
must be  negative semidefinite in order to model exponential
attenuation in passive media. Finally, one can also split
the Hamiltonian into the sum of Hermitian and anti-Hermitian parts.   

After choosing a suitable split (\ref{split})
we make
a substitution $\Psi(t) = \exp(t{\cal V})\Phi(t)$
in the original
evolution equation (\ref{9}). The 
new wave function $\Phi(t)$ satisfies the equation
with a time dependent Hamiltonian
\be
\label{scht}
\dot{\Phi}(t) = e^{-t{\cal V}}{\cal H}_0e^{t{\cal V}}\Phi(t) \equiv
{\cal H}(t)\Phi(t)\ ,
\ee
to be solved with the same initial condition $\Phi(0)=\Psi(0)$. Applying 
the leapfrog method to (\ref{scht})   we get 
$\Phi(t+\Delta t) = \Phi(t-\Delta t) + 2\Delta t{\cal H}(t)\Phi(t)$
valid up to $O(\Delta t^3)$. Returning to the initial variables,
we arrive at the following recurrence relation
\be
\label{lfnew}
\Psi(t +\Delta t) = {\cal L}(2\Delta t)\Psi(t-\Delta t) +
2\Delta t {\cal L}(\Delta t) {\cal H}_0 \Psi(t)\ ,
\ee
where ${\cal L}(\Delta t) = \exp(\Delta t {\cal V})$.
The amplification matrix, $\Psi(t+\Delta t)=
{\cal G}_{\cal L}(\Delta t)\Psi(t)$, for the recurrence (\ref{lfnew})
satisfies the equation
\be
\label{lfam}
{\cal G}_{\cal L}(\Delta t) = {\cal L}(2\Delta t){\cal G}^{-1}_{\cal L}(\Delta t)
+ 2\Delta t {\cal L}(\Delta t){\cal H}_0\ .
\ee
A deviation of 
the approximate solution ${\cal G}_{\cal L}^n(\Delta t)\Psi(0)$
from the exact solution relative to the energy norm
is of order $\Delta t^2 $ for any $n>0$. Thus, the scheme
is convergent and, hence,
a conditional stability exists for a sufficiently
small $\Delta t >0$ according to a general theorem
of Kantorovich \cite{richt} that establishes a general 
equivalence between convergence and conditional stability.
The conditional stability of (\ref{lfnew}) 
can be understood from the following observation. 
 Solving (\ref{lfam}) by perturbation theory
in $\Delta t$, it is not hard to find that
\be
\label{glgv}
{\cal G}_{\cal L}(\Delta t) - {\cal G}_V(\Delta t)
= \Delta t^3 {\cal K}(\Delta t)\ , \ \ \ \ 
{\cal G}_V(\Delta t) = {\cal L}(\Delta t/2) 
{\cal G}_0(\Delta t){\cal L}(\Delta t/2)\ ,
\ee
where ${\cal K}(\Delta t)$ is regular in the vicinity
of $\Delta t=0$ and vanishes whenever ${\cal H}_0$ and
${\cal V}$ commute, and ${\cal G}_0(\Delta t)$ is the
amplification matrix when ${\cal V}$ is set to zero.
The von Neumann condition is satisfied for 
${\cal G}_0(\Delta t)$ for a sufficiently small
$\Delta t >0$. Hence, powers
of ${\cal G}_0(\Delta t)$ applied to $\Psi(0)$ cannot
produce any exponential norm growth. Powers
of ${\cal G}_V(\Delta t)$ differ from those of
${\cal G}_0(\Delta t)$ by factors that are  
powers of $e^{\Delta t {\cal V}}$ and, hence,
can only produce exponential attenuation of the 
amplitude. Indeed, let $\Psi_V(t) = e^{t{\cal V}}\Psi(0)$. Then 
$\partial/\partial t (\Psi_V,\Psi_V) = 2(\Psi_V, {\cal V}\Psi_V)
\leq 0$ since ${\cal V}$ is negative semidefinite.
Thus, the approximate solution produced by the amplification matrix
${\cal G}_V(\Delta t)$ has no exponential growth, while
differing, relative to the energy norm, from that
produced by ${\cal G}_{\cal L}(\Delta t)$ by order of $O(\Delta t^2)$.
Therefore the modified
leapfrog scheme can be made conditionally stable and 
as accurate as desired  by reducing the time step. 

It should be noted that our arguments do not
prove that there cannot be any exponential
growth of the norm $\|\Psi(t)\|=(\Psi(t),\Psi(t))^{1/2}$
in the modified leapfrog scheme (\ref{lfnew}). 
All we can claim is that $\|\Psi(t)\|\leq \exp(Kt)$
which is also true for the conventional leapfrog scheme.
The difference is that in the modified leapfrog scheme
$K\sim O(\Delta t^2) $ as we have argued (a consequence 
of (\ref{glgv}) and uniform boundedness of powers of 
${\cal G}_V$), while 
in the conventional leapfrog scheme the constant $K$ is 
independent of $\Delta t$. Hence a possible exponential
growth cannot be suppressed by reducing 
the time step in (\ref{lf1}),
while it can be done in (\ref{lfnew}).
 
\subsection{An example of the Lorentz model}

To illustrate our general method we give an example of
the Lorentz model commonly used to describe dispersive media.
In the field representation of the Hamiltonian for the
Lorentz model, we make the following decomposition
\be
\label{lfsplit1}
{\cal H}^F = \pmatrix{{\cal H}_0 & {\cal V}_{FM}\cr {\cal V}_{MF} &0}
+ \pmatrix{0&0\cr 0&{\cal H}_M^F} \equiv {\cal H}_{0}^F + {\cal V}^F\ .
\ee
Substituting this decomposition into (\ref{lfnew}) we arrive at the following scheme
\ba
\label{lflm1}
\psi^F(t+\Delta t) &=& \psi^F(t-\Delta t) + 2\Delta t {\cal H}_0\psi^F(t)  
+2\Delta t \sum_a {\cal V}_{FMa}\xi^a(t)\ ,\\
\label{lflm2}
\xi^a(t+\Delta t) &=& e^{2\Delta t\, {\cal H}_{Ma}^{F}}\ \xi^a(t-\Delta t) +
 2\Delta t\ e^{\Delta t\, {\cal H}_{Ma}^{F}}\ {\cal V}_{MFa}\ \psi^{F}(t) \ ,\\
\label{eHM}
e^{t{\cal H}_{Ma}^{F}} &=& e^{-\gamma_at}\left[\cosh \tilde{\nu}_a t +
\frac{\sinh\tilde{\nu}_at}{\tilde{\nu}_a}\left({\cal H}_{Ma}^{F} +\gamma_a
\right)\right]\ ,
\ea
where $\tilde{\nu}_a = (\gamma_a^2 - \omega_a^2)^{1/2}$ and
the six-dimensional columns $\xi_a$ are introduced in Section 3.
The exponential (\ref{eHM}) is easy to compute 
by expanding ${\cal H}_{Ma}^{F}$ in the Pauli matrix basis
(a basis for the Lie algebra $su(2)$)
and then by using the well known 
formula for the exponential of a linear combination of Pauli matrices.
For small attenuation, $\gamma_a < \omega_a$, we get $\tilde{\nu}_a =
i\nu_a$. The hyperbolic
functions in (\ref{eHM}) become the trigonometric ones and
$\tilde{\nu}_a$ is replaced by $\nu_a$. 
Eigenvalues of the matter Hamiltonian are $\lambda_a = -\gamma_a
\pm \tilde{\nu}_ a$. Hence, ${\rm
Re}\, \lambda_a < 0$ and amplitudes of the matter fields are always
exponentially attenuated as $t\rightarrow \infty$, unless $\gamma_a
=0$ leading to ${\rm Re}\, \lambda_a =0$. 

The stability is ensured if ${\cal H}_0^F$ satisfies the von Neumann
condition. 
Let $k_{max}$
be the maximal norm of all wave vectors of the initial wave packet
and $\omega_p^{max}$ be the maximal value of $\omega_p
=(\sum_a \omega_{pa}^2)^{1/2}$ as a function
of position, then a sufficient condition for stability reads
\be
\label{sclflm}
\Delta t\sqrt{c^2k^2_{max} +(\omega_p^{max})^2} \leq 1\ .
\ee
Here the idea of the frozen coefficients \cite{shin} has been used
again.
The left hand side of inequality (\ref{sclflm}) is nothing but
the spectral radius of $\Delta t{\cal H}^F_0$ with frozen plasma
frequencies so that $\omega_p = \omega_p^{max}$. Note that
it is not difficult to solve the characteristic 
equation for ${\cal H}_0^F$ with frozen plasma frequencies
by using the Fourier basis.
The scheme (\ref{lfnew}) becomes especially simple in the case of
small attenuation $\gamma_a < \omega_a$. In the complex
representation of the auxiliary fields (\ref{S}) (cf. (\ref{caux}))
the matter Hamiltonians ${\cal H}_{Ma}^F$ are diagonal and
the action of its exponential is reduced to multiplication by
a complex number $e^{i\nu_a\Delta t}$.

Finally, it should be mentioned that,
by rearranging operators in the split, namely, by moving
${\cal V}_{FM}$ and ${\cal V}_{MF}$ to ${\cal V}^F$ in
(\ref{lfsplit1}), 
 the stability condition
(\ref{sclflm})
can be weakened to  $\Delta t ck_{max} \leq 1$.
This would come at the price of having a more complicated 
expression for ${\cal L}(\Delta t)$. In the case of the Lorentz
model it can still be computed analytically.
The new split can also be viewed as the use of the induction
representation in the modified leapfrog scheme, ${\cal H}^I
={\cal H}_0^I + {\cal V}^I$ where ${\cal H}_0^I$ contains 
only the blocks of ${\cal H}^I$ with the $\mbf{\nabla}$ operator.
The proof of the weaker stability condition can be found in
\cite{maxwell}.

\section{ Metal gratings in the Drude formalism}
\setcounter{equation}0

Here we apply our method to gratings made of a
metal whose optical properties are 
described by the Drude formalism. This is an actual
numerical scheme used in simulations in Section 7. 
The metal dielectric constant as a function of frequency 
is given by
$$
\varepsilon (\omega )=1+\tilde{\chi}(\omega)=1-\frac{\omega
_{p}^{2}}{\omega (\omega +i\eta)}\ ,
$$ 
where $\omega _{p}$ is the plasma
frequency, which is zero in the vacuum and constant in the 
region occupied by the metal as shown in Fig.1,
and $\eta $ is the absorption. The model coincides with
a one-resonant Lorentz model if $\omega_0=0$ and $\eta
\equiv 2\gamma$.
To satisfy the Gauss law exactly in simulations, we use
the induction representation according to Section 5.II.
An auxiliary field is chosen so that its first component
equals ${\bf P}$ and the second is denoted $\mbf{\xi}$.
The Hamiltonian evolution equations are taken in the form
\ba
\nonumber
\dot{\bf D}&=& c\mbf{\nabla}\times {\bf B}\ ,\\
\nonumber
\dot{\bf B}&=&-c\mbf{\nabla}\times\left({\bf D}-{\bf P}\right)\ ,\\
\nonumber
\dot{\bf P}&=&\eta \mbf{\xi}\ ,\\
\nonumber
\dot{\mbf{\xi}}&=&-\eta\mbf{\xi}-\frac{\omega_p^2}{\eta}\ \left(
{\bf D}-{\bf P}\right)\ .
\ea
To apply the modified leapfrog scheme the Hamiltonian is split
into the sum
\ba
\label{splitd}
{\cal H}^I &=& \pmatrix{0&c\mbf{\nabla}\times & 0&0\cr
-c\mbf{\nabla}\times &0&c\mbf{\nabla}\times &0\cr
0&0&0& -\eta\cr -\omega_p^2\eta^{-1}&0&
\omega_p^2\eta^{-1}& -\eta}=
{\cal H}^I_0+{\cal V}^I\ ,\\
\nonumber  
{\cal V}^I &=&{\sf diag}(0,\ 0,\ 0,\ -\eta)\ .
\ea
Clearly, ${\cal V}^I$ is negative semidefinite because
$\eta >0$. The stability of the modified leapfrog scheme
requires that eigenvalues of ${\cal H}_0^I$ have zero 
real parts. This is indeed the case. In the Fourier basis,
the $\mbf{\nabla}$ operator becomes $i{\bf k}$. It is straightforward
to find the characteristic polynomial $\det({\cal H}_0^I-\lambda)$. Its nonzero
roots are $\lambda = \pm i\sqrt{c^2{\bf k}^2 +\omega_p^2}$. Hence 
the scheme is stable if the time step is chosen
so that the condition (\ref{sclflm}) is satisfied. In particular,
$k_{\max}$ can be set to $k_0$ being the step of the reciprocal
lattice and $\omega_p^{max}$ is the plasma frequency of the metal
(silver in our simulations, see Section 7). An explicit form 
of the modified leapfrog scheme for the split (\ref{splitd})
reads
\ba
\nonumber
{\bf D}(t+\Delta t) &=& {\bf D}(t-\Delta t) + 2c\Delta t
\mbf{\nabla}\times {\bf B}(t)\ ,\\
\nonumber
{\bf B}(t+\Delta t) &=& {\bf B}(t-\Delta t) - 2c\Delta t
\mbf{\nabla}\times \left[{\bf D}(t)-{\bf P}(t)\right]\ ,\\
\nonumber
{\bf P}(t+\Delta t)&=&{\bf P}(t-\Delta t) -2\eta\Delta t
\mbf{\xi}(t)\ ,\\
\nonumber
\mbf{\xi}(t+\Delta t)&=& e^{-2\eta \Delta t}\, \mbf{\xi}(t-\Delta t) 
-2\Delta t\omega_p^2\eta^{-1}\, e^{-\eta\Delta t}\,\left[{\bf D}(t)-{\bf
    P}(t)\right]\ .
\ea
The action of the curl is computed by the fast Fourier method
in combination with a change of variables that enhances the
sampling efficiency near the metal-vacuum interface. The details 
are in Section 7.

\section{Results for extraordinary transmission gratings}
\setcounter{equation}0

To test our method we applied it to study transmission properties of
metal and dielectric gratings suspended in vacuum. Transmission
and reflection
gratings have been subject of a number of
experimental and theoretical works \cite{treacy}--\cite{11b}, \cite{gr4}. 
The interest is stimulated by nearly
100\% transmission or reflection within narrow wavelength range with
possibility of using such grating as efficient filters. Moreover,
extraordinary optical transmission has been observed in the $2D$
hole arrays \cite{gr1}, further stimulating  theoretical and experimental interest to
transmission properties of nanostructured materials \cite{gr3}.

In Fig.1 we schematically represent our system comprising a metallic or
dielectric slab with gratings. $D$ is a grating period, $a$ the size of the
grating and $h$ its thickness. For the sake of
comparison with previously published results we have chosen $D=1.75\mu
m$ and $a=0.30\mu m$. Transmission and reflection coefficients are computed as
a function of the grating thickness $h$. An incident
electromagnetic wave packet propagates along the $z$-direction, 
normal to the slab. The polarization is such that the electric field vector is
perpendicular to the gratings, while the magnetic field is parallel to
them (the so called $p$
-polarization). Calculations are performed in a finite $(x,z)$ box of the
size -$15D<z<12D$, $-D/2<x<D/2$. A uniform mesh of typically $256$ knots is
used in the $x$-direction and a nonuniform mesh of $512$ knots generated by
the change of variables is used in the $z$-direction. The change of
variables is used to enhance the sampling efficiency near the two
interfaces in the $z$-direction.
Periodic boundary conditions are
insured in $x$ through the pseudospectral approach based on the Fast Fourier
Transform. To suppress an artificial reflection of the wave packet, absorbing
layers are introduced at the box boundaries $z=\pm 15D$. 
The initial wave packet is Gaussian,
which allows us
to obtain transmission (reflection) coefficients within the entire
frequency bandwidth of the initial wave packet by a {\it single} 
simulation. In what follows
we are mainly interested in transmission (reflection) of the radiation
with wavelengths larger than the grating period (zero order diffraction).
Thus, reflected or transmitted waves propagate in the direction of the
$z$-axis, the same as the incident radiation. Note, however, that this is not
a limitation of our method which allows for a priori extraction of the entire
scattering matrix for all wave vectors.

In Fig.2 we show transmission and reflection coefficients obtained for
metallic gratings of variable depth. The dielectric response of the
metal is described within the Drude formalism (see Section 6). 
Following \cite{gr2} we use $\omega
_{p}=9eV$ and $\eta =0.1eV$ representative for silver. As clearly seen in
the Figure, the transmission coefficients exhibit narrow resonances for
certain wavelengths. With increasing thickness of the gratings,
$h$, the  number
of resonant structures increases. Our results are in a full agreement
with previously published theoretical studies \cite{gr2}. The only difference
being, that the model used in \cite{gr2} assumes perfect metal
surfaces inside the gratings, neglecting possible absorption of the
radiation. This leads to $100\%$ transmission at the resonant frequencies.
In our case, a part of radiation is absorbed by the metal so that the
transmission never reaches $100\%$. 

The observed structures in transmission are
associated with resonant modes of the electromagnetic field produced by 
coupled surface electromagnetic modes (called surface plasmon polaritons)
and waveguide modes inside the gratings. Some of these resonances posses
relatively long lifetimes. This can be immediately inferred from their
width, as, e.g., the resonance located at $\lambda =1.1D$ in the case of the
grating with thickness $h=1.4\mu m$. Another way to observe the trapped
field resonances is to look at the time dependence of the field transmitted
in the $z$-direction. The signal in Fig. 3 is registered by a detector
placed at the distance of $3.5D$ behind the gratings. As seen in the figure,
as soon as the resonances are populated by an incident $25fs$-long pulse, they
radiate the field during at least $125fs$. Here the radiation time is
determined by the lifetime of the narrowest resonance located at $\lambda
=1.1D$. It is worth mentioning that, as shows the sum of the transmission
and reflection coefficients, the total absorption is largest at the
resonance positions, i.e., when the interaction time between the radiation
(trapped mode) and the metal is large. 

Finally, Fig. 4 shows the field
structure of a trapped mode corresponding to the narrowest resonance
observed with the $h=2.4\mu m$ thickness grating. The $E_{x}$ component of
the electric field is presented. It is obtained via sufficiently long time
propagation so that contributions from  less long-lived states
vanish.

In Figs. 5 and 6 we show reflection coefficients for dielectric gratings
suspended in vacuum. Simulations are performed using the conventional
leapfrog scheme (\ref{lf1}) applied to (\ref{schr}). Since the
attenuation is absent, the scheme is stable if the time step
satisfies (\ref{scdiel}).
The dielectric material is modeled through the
frequency independent dielectric constant $\varepsilon =2$ (Fig.5) and $
\varepsilon =4$ (Fig.6). Only the reflection coefficient is shown here since
there is no absorption of the radiation in dielectric so that the
transmission can simply be inferred from the unitarity of the scattering
matrix. Without gratings the dielectric slabs are basically transparent in
both cases. Introducing grating structures results in a complete
reflection of the incident radiation within an extremely narrow wavelength
bandwidth. The associated guided mode resonances have been extensively discussed
in the literature \cite{gr4}. In the case of $\varepsilon =2$ the resonances are so
narrow that extraction of frequency dependent transmission and
reflection coefficients by the Fourier transform of the scattered
wave would require too large propagation time (see also
Fig. 7). We had to stop our wave packet propagation before the
radiation emitted by
resonances ended. This explains why the reflection
coefficients do not reach its maximal value $1$ in this case. Consistently with
the metal case, the number of resonant structures increases with increase
of the width $h$ of the gratings. The width of resonances increases with $%
\varepsilon $ as follows from the comparison of Figs. 5 and 6. Note also
that resonances are associated with Fano profiles that usually arise because
of the interference between the non-resonant and resonant contributions to
the scattered wave. Such narrow reflection structures in the case of
dielectric gratings have been usually studied by 
stationary methods in the frequency domain. An important 
advantage of the time-dependent
study is that one has an immediate access to all the details 
the temporal evolution of electromagnetic fields in any desired
part of the system. 

In Fig. 7 we show the time dependence of the field transmitted in
the $z$-direction for gratings characterized by $\varepsilon =2$ and $%
h=0.8\mu m$. With these parameters there is only one resonance in the reflection
spectrum. The signal is registered by a detector placed at the distance of $%
3.5D$ behind the gratings. First, we observe that a $25fs$ incident pulse 
is transmitted through the structure without modification. 
The lasing effect, when the transmitted field is followed by basically
monochromatic radiation, can clearly be seen.
For readability reasons we could not show the complete time evolution in the
figure, but the lasing effect lasts for at least (!) $2ps$ reflecting
an extraordinary long lifetime of the trapped resonant field. It is this radiation
which comes in the phase opposite to the corresponding harmonic in the transmitted
initial pulse and leads, finally, to the zero transmission at the corresponding
frequency. The same lasing effect to the left from the grating structure
is responsible for the $100\%$ reflection at the same frequency. 

In Fig. 8 we show a
typical structure of the field corresponding to a trapped (resonant) mode.
The $D_{x}$ component of the electric induction is represented in the case of
dielectric gratings with $\varepsilon =4.$ The thickness of
the dielectric slab is $h=0.6\mu m$. Note that in contrast to the  
metal grating structure,
the field in the present case occupies the entire slab and not only the
vacuum part of the grating.

\section{Conclusions}

We have developed a time domain algorithm for the initial value
problem for the Maxwell's theory of linear passive media.
The algorithm is based on ({\it i}) the Hamiltonian formalism for
evolution differential equations, ({\it ii}) the Fourier 
pseudospectral method in which the sampling efficiency in designated 
space regions is enhanced by a suitable change of variables, and
({\it iii}) the modified leapfrog scheme. We have analyzed
the stability of the algorithm and found explicit stability 
conditions when passive media are described by multi-resonant
Lorentz models. We have implemented and tested our algorithm
for extraordinary transmission and reflection gratings whose
optical properties have been studied in a number of theoretical
and experimental works. Numerical simulations based on our algorithm 
are shown to produce extremely accurate
data for the well studied far-field (zero-order diffraction) 
at relatively low computational
costs. A single simulation of an incident wideband wave packet  
is sufficient to determine transmission and reflection 
properties of the gratings in the frequency range 
of the initial wave packet.
In addition, our algorithm allows us to see a real
time dynamics of formation of long-living resonant excitations
of electromagnetic fields in the grating as well as formation
of transmitted and/or reflected wave fronts in the entire
frequency range of the initial wave packet. 
It is believed that our algorithm would be useful for numerical studies
of other nanostructured materials.

\vskip 1cm
\noindent
{\Large\bf Acknowledgments}

S.V.S thanks the Laboratory of Atomic and Molecular Collisions
of the University of Paris-Sud for the warm hospitality and
Dr. R. Albanese (Brooks, Air Force Base, TX) for his continued
interest and support of this project.
The work of S.V.S. has been supported in part by US Air Force
Grants F4920-03-10414 and F49620-01-1-0473.

\newpage

\noindent
{\large \bf Figures}

Fig. 1 A schematic representation of the studied system. The incident
wave packet propagates 
along the normal to the slab containing the gratings (z-direction). 

Fig. 2. Calculated zero-order transmission (solid lines) and reflection
(dashed lines) coefficients for metallic gratings described in the text.
The results are presented as a function of the wavelength $\lambda $ of the
incident radiation measured in the units of the period of the gratings, $D$.
Different panels of the figure correspond to the different thickness $h$ of
the gratings, as indicated.

Fig. 3. The electric field measured by a  detector placed behind the metallic
gratings with thickness $h=1.4\mu m$. Only the field corresponding to the
zero-order transmitted wave propagating along the $z$-axis is represented.
It is obtained by the Fourier analysis of the $x$-coordinate dependence of
the field at the detector position. The signal is shown as a function of
time measured in femtoseconds.

Fig. 4. A snapshot of the $x$-component of the electric field, $E_x$ for the 
grating thickness $h=2.4\mu m$. The results are presented as a function of $x$
and $z$ coordinates measured in units of $D$. 
Red and blue colors correspond, respectively, to positive and negative values of the
field. The snapshot has been produced after a sufficiently long
propagation time so that the field pictured in it indeed
corresponds to the long-lived resonance giving enhanced transmission
at $\lambda =1.15D$ (see Fig.2).

Fig. 5. Calculated zero-order reflection coefficient for dielectric gratings
with $\varepsilon =2$. The results are presented as a function of the wavelength $%
\lambda $ of the incident radiation measured in the units of the period of
the gratings, $D$. Different panels of the figure correspond to the
different thickness $h$ of the gratings, as indicated.

Fig. 6. Calculated zero-order reflection coefficient for dielectric gratings
with $\varepsilon =4$. Results are presented as a function of the wavelength 
$\lambda $ of the incident radiation measured in the units of the period of
the gratings, $D$. Different panels of the figure correspond to the
different thickness $h$ of the gratings, as indicated.

Fig. 7. The electric field measured by a  detector placed behind the dielectric
grating with thickness $h=0.8\mu m$ and dielectric constant
$\varepsilon =4$. 
Only the field corresponding to the zero-order transmitted wave
propagating along the $z$-axis is represented. It is obtained by the
Fourier analysis of the $x$-coordinate dependence of the field at the
detector position. The signal is shown as a function of time measured in
femtoseconds.

Fig. 8. A snapshot of the $x$-component of the electric induction, $D_x$ for
the dielectric grating with dielectric constant $\varepsilon =4$ and
thickness $h=0.6\mu m$. The results are presented as a function of $x$ and $z$
coordinates measured in units of $D$. Red and blue colors correspond, 
respectively, to positive and negative
values of the induction. The
snapshot has been produced after a sufficiently long propagation time
so that the induction pictured in it indeed
corresponds to the resonance giving enhanced transmission at $\lambda \cong
1.3D$ (see Fig.6).

\end{document}